\documentclass[]{spie}  
\usepackage[dvips]{graphicx}

\title{The RAPTOR Experiment: A System for Monitoring the Optical Sky in Real Time} 

\author{W. T. Vestrand, K. Borozdin, S. Brumby, D. Casperson, E. Fenimore,
M. Galassi, \\ 
K. McGowan, S. Perkins, W. Priedhorsky, D. Starr, R. White,
P. Wozniak, and J. Wren\\
\skiplinehalf
Los Alamos National Laboratory, Los Alamos, NM, USA \\}

\authorinfo{Further author information: (Send correspondence to W.T.V.)\\W.T.V.: E-mail: vestrand@lanl.gov, Telephone: 1 505 665 9542\\   Address: Los Alamos National Laboratory, MS-D436, Los Alamos, NM 87545, USA}

\begin{document} 
\maketitle 

\begin{abstract}
The Rapid Telescopes for Optical Response (RAPTOR) experiment is a spatially distributed system of autonomous robotic telescopes that is designed to monitor the sky for optical transients. The core of the system is composed of two telescope arrays, separated by 38 kilometers, that stereoscopically view the same 1500 square-degree field with a wide-field imaging array and a central 4 square-degree field with a more sensitive narrow-field ``fovea" imager.  Coupled to each telescope array is a real-time data analysis pipeline that is designed to identify interesting transients on timescales of seconds and, when a celestial transient is identified, to command the rapidly slewing robotic mounts to point the narrow-field ``fovea'' imagers at the transient. The two narrow-field telescopes then image the transient with higher spatial resolution and at a faster cadence to gather light curve information. Each ``fovea" camera also images the transient through a different filter to provide color information.  This stereoscopic monitoring array is supplemented by a rapidly slewing telescope with a low resolution spectrograph for follow-up observations of transients and a sky patrol telescope that nightly monitors about 10,000 square-degrees for variations, with timescales of a day or longer, to a depth about 100 times fainter. In addition to searching for fast transients, we will use the data stream from RAPTOR as a real-time sentinel for recognizing important variations in known sources. All of the data will be publically released through a virtual observatory called SkyDOT (Sky Database for Objects in the Time Domain) that we are developing for studying variability of the optical sky. Altogether, the RAPTOR project aims to  construct a new type of system for discovery in optical astronomy---one that explores the time domain by ``mining the sky in real time". \end{abstract}



\section{INTRODUCTION}
\label{sect:intro}  

While it has been known for centuries that the optical sky is variable, monitoring the sky for optical transients with durations of less than a day is a rich area of research that remains largely unexplored.\cite{Paczynski00,Nemiroff99} The fact that spectacular celestial transients exist was clearly demonstrated by the detection \cite{Akerlof99}, with the Robotic Optical Transient Search Experiment I (ROTSE-I) telescope located at Los Alamos National Laboratory (LANL), of an optical transient associated with a Gamma Ray Burst  (GRB) at redshift z=1.6. The optical flash generated by that cosmological explosion lasted about 80 seconds and reached an astounding peak apparent magnitude of 9, making it the most luminous optical source ever measured by man ($M_V=-36.4$).  However, without the real-time position provided by a high-energy satellite that cued robotic optical telescopes to slew to the correct position, the remarkable transient, that was observable potentially even with binoculars, would have been missed.

 There are also reasons to suspect the existence of celestial optical transients that cannot be found through sky monitoring by high-energy satellites. Theoretical models of GRBs have suggested that so-called optical ``orphan" transients might be a thousand times more common than the GRBs\cite{Granot02}
 and/or that optical transients could be precursors to GRBs\cite{Paczynski02}. Further, our knowledge of the variability of optical sky is so incomplete that it is likely that there are new, as yet undiscovered, classes of rapid optical transients that are completely unrelated to high-energy transients. However, to search for optical transients, astronomers need a wide-field optical monitoring system that can autonomously locate, in real time, celestial transients with timescales as short as a minute. Such a system has never been constructed.

Beyond the study of optical transients with timescales of minutes, there is a long and diverse list of other interesting scientific subjects that need wide-field optical monitoring for efficient exploration.\cite{Paczynski97,Paczynski01b}. The subjects range from searches for killer asteroids and extrasolar planetary systems, to eclipsing and pulsating stars, to novae and supernovae, to large amplitude outbursts from active galactic nuclei. Once found, those rare objects or fleeting events are best studied with large, narrow-field, telescopes---but those powerful instruments have very limited capability to find the interesting targets. Wide-field optical monitors can therefore enable otherwise impossible observations.

The idea of monitoring the dynamic optical sky is an old one, but it is only with 21st century information technology that one can seriously contemplate a systematic monitoring of all the sky for optical transients with durations as short as a minute. For ground-based telescopes, the optical sky contains half a trillion optical resolution elements. Monitoring the entire visible sky, with that resolution, for transients that last only a minute will generate a continuous data rate approaching 10 Gigabytes per second. To be effective, that huge data stream needs to be mined in real time to locate the transient and cue follow-up observations. Everyone agrees that such a system is likely to make important discoveries, but clearly the construction of such a high-resolution system that monitors the full sky, all of the time, is challenging. Here we discuss a program underway at Los Alamos National Laboratory that is taking the first steps toward that goal.

\section{The RAPTOR system concept: An analogue of Human Vision}
\label{sect:intro}  

The RAPid Telescopes for Optical Response (RAPTOR) program is a pathfinding effort to construct a fully automated sky monitoring system that locates, and begins follow-up studies in real time of, celestial optical transients with timescales as short as a minute.\cite{Vestrand01} The RAPTOR system is a distributed system of telescope arrays constructed mostly of modest commercial off-the-shelf (COTS) components, but employing a sophisticated software pipeline with a real time alert system. It makes sense to begin with such a system because the key challenges for real-time all-sky monitoring are the construction of a real-time pipeline and the development of techniques for identifying celestial transients in the ``forest'' of non-celestial transients. To fully exploit the potential for scientific discovery of the project, the RAPTOR sky monitoring system is a closed-loop system that autonomously follows up the optical transient detection, within seconds, with more sensitive telescopes to verify the detection and to measure the spectrum and temporal variations of the transient.  Another key goal of the project is to develop data mining and machine learning tools that will allow us to use the data stream from the sky monitoring system as a real-time sentinel for the recognition important variations in known sources. The project is also constructing a virtual observatory for studying the variability of optical sky, SkyDOT (Sky Database for Objects in the Time Domain), that will publicly release all the RAPTOR data. In short, the RAPTOR project aims to use aggressively advanced information technology techniques and robotic hardware to construct a system for sky monitoring that allows one, for the first time, to mine the optical sky in real time.

 As predators, we humans have evolved a highly sophisticated vision system for both imaging and change detection.\cite{Hubel95} Human vision employs two spatially separated eyes viewing the same scene both to eliminate image faults like ``floaters" and to extract distance information about objects in the scene. Each eye has a wide-field, low-resolution, imager (rod cells of the retina) as well as a narrow-field, high-resolution imager (cone cells of the fovea). Both eyes send image information to a powerful real-time processor, the brain, running ``software" for the detection of interesting targets. If a target is identified, both eyes are rapidly slewed to place the target on the central fovea imager for detailed ``follow-up" observations with color sensitivity and higher spatial resolution. During each step of the process, our brain is running powerful real-time software and comparing with an adaptive catalog---our memory---to identify and study changes in the scene.

The RAPTOR system concept is an analogue of human vision. The subsystem for rapid transient identification employs two, spatially separated, telescope arrays (RAPTOR-A and RAPTOR-B). Each telescope array simultaneously images the same 1500 square-degree field with a wide-field imager and a central 4 square-degree with a narrow-field ``fovea" imager. The real-time software pipeline instantly analyzes images from RAPTOR A and B, potential candidates are identified, and the positions of any interesting transients without a measurable parallax are fed back to the mount controllers with instructions to point the fovea telescopes at the transient. The two fovea cameras then image the transient with higher spatial resolution and at a faster cadence to gather light curve information. Each fovea camera also images the transient through a different filter to provide color information. The RAPTOR A and B arrays therefore act as a binocular monitoring system employing closed loop feedback that autonomously identifies, generates alerts, and makes detailed follow-up observations of optical transients in real-time. This sky monitoring array is supplemented by a rapidly slewing telescope with a low resolution spectrograph  (RAPTOR-S) for follow-up observations of transients and a sky patrol telescope (RAPTOR-P) with a field-of-view of 60 square-degrees that nightly monitors about 10,000 square-degrees for variations, with timescales of a day or longer, to depth about 100 times fainter than the stereoscopic system.

\subsection{RAPTOR System Hardware}
\subsubsection{The Stereoscopic Wide-Field Monitoring System: RAPTOR-A and RAPTOR-B} 
\label{sect:ide-field}

Previous attempts to search the sky for rapid optical transients from celestial objects have always been compromised by false triggers\cite{Vanderspek94,Kehoe02}. Those non-celestial false triggers are generated by a wide range of noise sources including---but not limited to---cosmic-ray hits, hot pixels, aircraft lights, image ghosts from bright stars, meteors, and glints from space debris as well as satellites. One observing strategy that has been used to reject many of the false positives is to require detection of the transient in consecutive images. This persistence requirement filters out many false positives, but glints
from slowly moving objects, like those in geostationary stationary orbit, can still be falsely identified as celestial. In addition, one loses all ability to find even bright flashes that have a duration shorter than the single image integration time.

To suppress false triggers, the RAPTOR system uses two wide-field arrays (RAPTOR-A and RAPTOR-B) to stereoscopically view the same scene. The RAPTOR-B telescope is located at the Los Alamos Neutron Science Center (LANSCE) in Los Alamos, New Mexico, and the RAPTOR-A telescope array is located almost 38 kilometers due west at our Fenton Hill Observatory site in the Jemez mountains. That 38 kilometer baseline yields a parallax shift of more than 220 arcseconds for non-celestial objects all the way out to the altitude of geostationary orbits at 36,000 kilometers. Our wide-field imagers have a single pixel resolution of 34 arcseconds so any transient generated at distance at least out to six times geostationary will have a detectable parallax. During the initial commissioning operation of the system we have imposed strict filters that require a transient to have no measurable parallax and appear in two consecutive 30-second exposure frames at both sites. However, ultimately we plan to relax the trigger constraint to allow follow-up of transients that show no parallax in simultaneous stereoscopic image pairs. That mode of operation will allow us to also search for bright celestial transients that have a duration shorter than our standard 30-second integration time.

The Wide-field imagers of the RAPTOR-A and RAPTOR-B arrays are each composed of four Canon 85mm f/1.2 lenses with CCD cameras at the focal planes (see figure 1). The cameras are thermo-electrically (TE) cooled Apogee AP-10 cameras, which employ a 2Kx2K format Thomson 7899M CCD chip with 14-micron pixels. The camera electronics have been optimized to provide fast readout of the entire array in 5 seconds. Each camera of the array covers a $19.5^\circ\times19.5^\circ$ field and the  four cameras are splayed so that they together mosaic a field of approximately 1500 square-degrees. The limiting magnitude of this wide-field system is $\sim12^{th}$ magnitude for a thirty-second exposure at the LANSCE site and $\sim12.5^{th}$ magnitude at the darker Fenton Hill site. While the sensitivity of this system is modest, the simple approach of mosaicing the field with photographic lenses and amateur-grade CCD cameras can be easily scaled to cover the entire sky. Further, since each element of the array has its own dedicated computer for control and data acquisition, the signal processing task is intrinsically parallel---which is a significant advantage when searching in real time for short duration transients. 

In the center of each wide-field array is a fovea telescope. It is composed of a large 400mm focal length Canon telephoto lens with a 5.6-inch objective diameter and a Finger Lakes Instruments (FLI) MaxCam CM2-1 CCD camera. The camera uses a TE cooled, 1Kx1K format, Marconi CCD-47 back-thinned chip with 13-micron pixels and a fast 4-second readout time. In this configuration the fovea cameras cover approximately a $2^\circ\times2^\circ$ field-of-view and have nearly five times the spatial resolution of the wide-field array. The limiting magnitude of these telescopes is about 16.5 magnitude for a 60-second exposure, making then well suited for faster cadence imaging of any transient identified by the wide-field arrays. To provide color information about the transient, the fovea camera for RAPTOR-A images through a Gunn r filter and the fovea camera for RAPTOR-B images through a Gunn g filter.

Each telescope array is mounted on a rapidly slewing robotic mount.  The mounts were designed to have the capability to place the fovea camera on the transient and to begin follow-up observations within a few seconds no matter where the transient occurred in the 1500 square-degree monitoring field. Tests of the completed mounts indicate that they can accelerate at rate of up to 400 degrees/sec$^2$ and reach speeds of 200 degrees/sec. In practice this means that the arrays can be slewed from horizon to horizon in 1.5 seconds and, after waiting another second for mount and telescope vibrations to damp, begin imaging. To our knowledge, the RAPTOR mounts are the swiftest ever constructed for astronomical purposes.

In routine operation, both telescope arrays will track a single 1500 square-degree field near zenith for one hour, from field center at a position 30 minutes before meridian crossing to 30 minutes after meridian crossing, before moving to a new field. This will generate about 100 simultaneous, 30-second duration, image pairs of a given field per night. Such fast cadence imaging will optimize our sensitivity for triggering on fast transients and will also allow us to minimize the systematic errors on our photometry for planetary transit searches.

   \begin{figure}
   \begin{center}
   \begin{tabular}{c}
   \includegraphics[height=10cm]{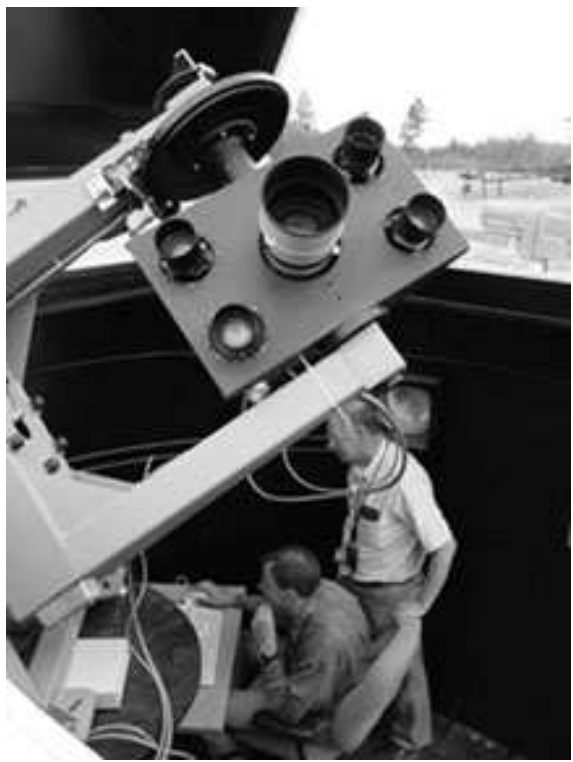}
   \end{tabular}
   \end{center}
   \caption[example] 
   { \label{fig:example} 
The RAPTOR-A telescope array which is part of the stereoscopic wide-field monitoring system. The four wide-field imaging cameras surround a central fovea telescope and the entire array is mounted on a rapidly slewing robotic mount. The telescope is housed in a robotic enclosure and is designed to run in a fully autonomous manner.}
   \end{figure} 

\subsubsection{The Spectroscopic Follow-Up System: RAPTOR-S} 
\label{sect:narrow-field}

The RAPTOR stereoscopic sky monitoring system will generate real-time alerts for distribution to more powerful follow-up instrumentation. These alerts will contain sufficiently accurate position (a few arcseconds from the fovea imagers) and brightness measurements to enable the best possible spectrographic and photometric observations by other instruments. Eventually, of course, we intend to distribute our alerts to the growing, worldwide network of telescopes that are used to follow-up triggers from satellites searching for gamma ray bursts. However, to prove the utility of the alerts and make the first studies of the physics of these optically selected transients, we intend to make spectroscopic follow-up observations with our own dedicated equipment.

We have therefore constructed the RAPTOR-S telescope as an autonomous robotic telescope for prompt follow-up spectroscopy of any transient detected by the sky monitoring system. The telescope is a 30-cm F7 Ritchey-Chretien telescope with a two-element field flattener that yields a spot size smaller 20 microns across the central 40mm diameter image circle. The nominal field-of-view for the telescope, when employing the full 40mm image circle, is 1.1 degrees. The ruggedized telescope incorporates Invar rods to minimize temperature induced focus variations and an interferometrically matched front window constructed of fused silica to seal the tube from dust and moisture. To obtain spectral dispersion, a transmission grating is located between the field flattener and a sensor-mounting flange that can accommodate either a conventional CCD camera or a  photon-counting imaging spectrophotometer.  This configuration, when used in first order with a grating of 300 grooves/mm, yields a spectral resolution of $R=\lambda/{\Delta\lambda}\sim200$ at 600 nm with an AP-6E CCD camera and has a sensitivity well matched, $\sim12^{th}$ magnitude for a 60-second integration, to the sensitivity limit of our stereoscopic sky monitoring system. The entire tube assembly is held by a rapidly slewing mount that is identical to those used by the RAPTOR-A and RAPTOR-B telescopes. Our goal is to point this spectrograph at the transient within seconds to obtain information about the shape of the continuum, the migration of the emission peak through the optical band, and search for spectral lines in the prompt emission.   As well as responding to alerts from our sky monitoring system, this system is connected to the GCN network\cite{Barthelmy98}
 so that it can respond to external gamma ray burst alerts.  
   \begin{figure}
   \begin{center}
   \begin{tabular}{c}
   \includegraphics[height=10cm]{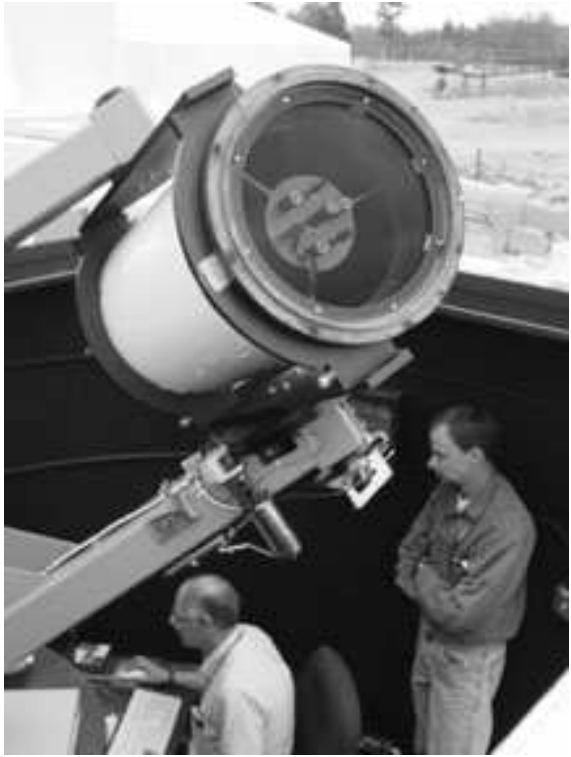}
   \end{tabular}
   \end{center}
   \caption[example] 
   { \label{fig:example} 
The RAPTOR-S telescope at Fenton Hill Observatory. It is a 0.3-m Ritchey-Chretien telescope with a low resolution spectrograph that responds in seconds to real time alerts generated by the wide-field sky monitoring system.}
   \end{figure} 

\subsubsection{The Sky Patrol System: RAPTOR-P}
\label{sect:sky patrol}
We are also constructing a sky patrol telescope array (RAPTOR-P) at the Fenton Hill Observatory site. This patrol telescope array will be composed of four Canon 400mm f2.8 telephoto lens that are identical to those used for the fovea telescopes of RAPTOR-A and RAPTOR-B. Each lens will be mounted to a large format 2Kx2K CCD camera with a filter wheel assembly to form an array that yields an instantaneous field-of-view of 60 square-degrees and a single pixel resolution of 6 arcseconds. The filter wheel for each camera will hold Gunn g and r filters that are identical to those employed by the fovae cameras of the stereoscopic sky monitoring system. Initially the array will use unfiltered Apogee AP-10 CCD cameras, but ultimately we plan to use cameras employing back-thinned Marconi CCD chips. With the back-thinned CCDs, the array will have a limiting unfiltered magnitude limmit of nearly $16.5$ magnitude for a 60-second integration. 

The primary goal of the RAPTOR-P observing program is the construction of an adaptive catalog that will be used to refine the accuracy of real-time event localization by the stereoscopic sky monitoring system. The transient will be localized to a accuracy of $\sim$30 arcseconds by the wide-field cameras, but to enable follow-up with narrow-field telescopes and high-resolution spectrographs, we would like to refine the real-time locations to an accuracy of a few arcseconds. Whenever a transient is recognized by the wide-field monitoring system, the mounts slew to point the higher spatial resolution fovea telescopes at the coarse position to ``zoom in" on the transient location. But to identify the transient in the fovea image in real time, and refine the positional accuracy to a few arcseconds, the system needs a reference catalog containing the locations of known sources with spatial resolution and depth comparable to that of the fovea image. The RAPTOR-P system is designed to provide that reference catalog and continuously update the catalog contents as sources vary---a so-called adaptive catalog.

RAPTOR-P will gather two color photometric observations for about 10,000 square-degrees per night. This will allow us to search for variations, with timescales of a day or longer, to depth about 100 times fainter than the wide-field cameras of RAPTOR-A and RAPTOR-B. Further, it will allow us to study the variability of nearly 30 million objects over the course of a year. As we briefly discuss in section 3, this will allow us to explore a diverse range of astronomical questions in the time domain. 

\subsection{The Real Time Analysis Pipeline}

As the analogy with Human vision suggests, constructing the real time analysis pipeline and control software---the brain---is one of the key challenges of the project. To keep pace with the RAPTOR data stream, the software pipeline must be able to make photometric and astrometric measurements for 250,000 objects and identify any new sources in less than 30 seconds. This is nearly an order of magnitude faster than the fastest existing real-time pipelines. To meet this challenge we had to engineer a new pipeline designed for speed and optimized for rapid recognition of transient sources. 

To achieve the required performance we employed an advanced type of architecture that, until now, has been rarely used for astronomical software. This architecture has a particular type of flexibility: one that provides every piece of functionality as library API (Application Programming Interface) functions. Those functions operate on memory-resident data or tight network data transfers. In the case of a single telescope and computer this allows a fast and tight monolithic pipeline. When many telescopes are combined, the API components distribute cleanly and efficiently to the various sites. We overhauled some existing astronomical software to fit our architecture (such as an overhaul of SExtractor, a standard source extraction program, to make it an embeddable library), but most of the pipeline components were designed and implemented from scratch.

The full real-time pipeline (shown in Figure 3) uses a diverse collection of components and algorithms, ranging from data acquisition to source extraction, astrometry, relative photometry corrections and the actual smarts of transient detection.  Immediately upon completion of each exposure, the raw images are combined with flat-field and dark frames to form corrected images. Sources are then extracted using our modified version of the SExtractor package to form a source-object file for each corrected image. Using a reference star catalog, the extracted source-object files are then registered to calculate the source coordinates and the relative photometry is derived to form a calibrated object file for each image. The entire process of calibrated list extraction is accelerated to take less than 10 seconds and runs in parallel for all ten of the stereoscopic array cameras.

Deriving the astrometric and photometric corrections for forming the calibrated object list is a challenge for extremely wide field images like those provided by our RAPTOR sky monitoring system. Large field distortions are common as well as large photometric systematic errors like  residual vignetting and actual sky brightness gradients that affect flat-fields, gradients in atmospheric extinction, and occasional partial cover by thin clouds. Starting with a list of detections and the reference catalog, our software uses a triangle search technique to match 25 of the brightest objects and derive a zeroth order coordinate transform. Higher order polynomial fits are then derived by matching the next tier of bright stars and decreasing the matching radius until good, spatial residuals $<0.1$pixel, fits are obtained. Typically the RAPTOR-A and RAPTOR-B images require 3rd order fits, while the narrower field ``fovea" images only require 2nd order polynomials to achieve accurate coordinate derivation. For relative photometric corrections, we find that accurate results (residuals of $<2\%$) are obtained by mapping the systematics with a grid of macro-pixels, each containing a region of about 100$\times$100 detector pixels, and bilinearly interpolating for any given location in the frame.

The next component of the real time analysis pipeline is the ``transient detection" software. The key problem for a practical real time sky monitoring system is the elimination of false triggers. To combat those false triggers, the RAPTOR telescope arrays are separated by 38 kilometers to enable stereoscopic viewing of the scene. Unfortunately, that large spatial separation imposes severe limits on the bandwidth for communication between the two arrays. So while image differencing has some advantages for identifying transients, bandwidth limitations imposed by the T-1 line communication forced us to employ an approach that compares the calibrated object lists derived at each site with a resident adaptive catalog. In its current form, the triggering software employs pair matching of object lists from consecutive images at a single site and then sends the short list of candidates to a central location and looks for a match in lists from the two different sites.\cite{Borozdin02} For a typical single image, matching with our internal adaptive catalog yields about 500 candidates compared to  3,000 when matching with a traditional catalog like the Hubble Guide Star Catalog. Pair matching the candidates in consecutive images yields about 100 candidates per exposure, but dithering the mount position between exposures reduces the false triggers to about 1 per exposure. When the pairs from both sites are combined without a signal to noise cut, we expect to achieve a false coincidence rate of one every 100 exposures or a few per hour. Most of those false triggers are at the limiting magnitude of the system, so a low signal-to-noise cut should push the false trigger rate to less than a few per night.

All of the real-time pipeline is finished and has been tested on real data. It was clocked at 18.5 seconds for the entire analysis process through transient identification. This exceeds our design goal of 30 seconds and is an order of magnitude faster than any other existing pipeline for transient recognition.

A final challenge for the RAPTOR software program has been the development of the software for data acquisition and management of the spatially distributed autonomous system. That software controls the observatory enclosures, telescope mounts, and cameras. Further, it synchronizes the operation telescope arrays separated by 38 km and, when an internal or external alert is received, manages prompt response in a follow-up mode of operation. Our solutions to those distributed control problems for the hardware are discussed in Ref.\citenum{Wren02}.
   \begin{figure}
   \begin{center}
   \begin{tabular}{c}
   \includegraphics[height=9cm]{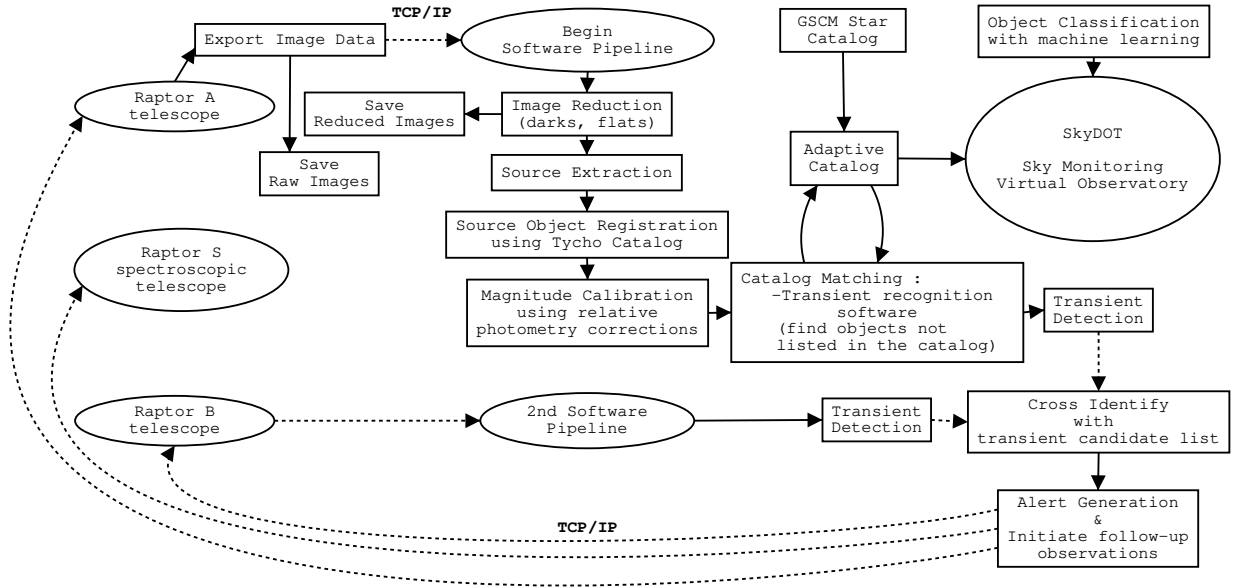}
   \end{tabular}
   \end{center}
   \caption[example] 
   { \label{fig:example} 
A flow chart showing the basic components of the RAPTOR real time pipeline and the feedback loop. }
   \end{figure} 

\subsection{A Virtual Observatory: SkyDOT}

The RAPTOR system will make two-color monitoring observations of nearly 30,000,000 celestial objects over the course of a year. To handle that enormous volume of data, as well as expedite public release of the data, we are constructing a Virtual Observatory called SkyDOT (Sky Database for Objects in the Time Domain)\cite{Wozniak02}. As the name implies, SkyDOT is dedicated to studying the time derivative, or variability, of celestial objects. An important feature of SkyDOT is that it will be updated in near real time so that users can promptly access the most recent measurements of a given object as well as its variability history from the website. The site will also provide a set of high level tools for data analysis and visualization.

One of the longer term goals of the RAPTOR project is to develop machine learning tools that can automatically classify the nature of the variability in all of the monitored objects and recognize when a given object has entered an "anomalous" state. If successful, that software will act a sky sentinel that continuously monitors the RAPTOR real time data stream and identifies objects that need follow-up with more powerful narrow-field telescopes. Such a real time notification system will enable otherwise impossible observations.

\section{RAPTOR Science}
\subsection{Optical Transients}

RAPTOR will search, unbiased by the triggering criteria of gamma ray detectors, for bright optical transients with timescales as short as a minute. To fully understand the physics of those optical transients, we also contructed a rapid response telescope that is dedicated to real time spectroscopic follow-up observations of the transients. When spectrographs on large telescopes were pointed at GRB positions, hours or days after the event, they found faint, fading optical counterparts with spectral features that indicate an origin at cosmological distances. Those spectral observations also taught us that GRBs are more amazing than almost everyone had dared to speculate---they are the largest explosions since the big bang and contribute in a substantial way to the entire energy output of the universe. Further, for the gamma-rays to escape the intense internal radiation fields requires the emitting material to be driven to Lorentz factors of 100 to 1000, that is, a velocity greater than 0.9999 the speed of light. Such extreme ultra-relativistic bulk flows occur nowhere else in the Universe. The extreme ultra-relativistic nature of the outflow means that the complex history of interactions and deceleration of the bulk flow, which occurs on the timescale of a day in the plasma frame, is carried by emission that arrives at Earth within the span of a few minutes. Therefore observations taken hours after the burst only measure emission from regions located light years away from the explosion. Our closed-loop RAPTOR system is designed to recognize optical transients in real time and in seconds begin follow-up spectroscopy observations while the transient is still bright and carrying the signatures of the cataclysmic event.  

\subsection{ Near Earth Objects and Killer Asteroids}

Every year a small asteroid 20 meters across collides with Earth \cite{Rabinowitz93,Stuart01}
 and several hundred miss by less than one quarter of the distance to the moon ($10^5$ km). In the night sky, those small asteroids would appear as objects that are brighter than the $\sim 12.5^{th}$ magnitude limit of the RAPTOR sky monitoring system. Traveling with typical velocities of $\sim 10$ km/sec they traverse the earth to moon distance in eight hours and, should they strike the earth's atmosphere, can generate an explosion that is the equivalent of a few Megatons of TNT. Moving nearly perpendicular to our line of sight at a distance of $10^5$ km, they would have a proper motion of 20 arcsec/sec or 2,000 RAPTOR pixels per hour. Of course the most interesting ones are heading directly at us and have a much smaller proper motion. But once they move within $10^5$ km of the earth they will have a detectable parallax even with the RAPTOR wide-field imagers and would appear as persistent objects with a measurable parallax. Less common, but more dangerous, are 100 meter rocks like the one that was recently discovered (June 2002) after it had missed earth collision by only a third of the distance to the moon. If it had collided, it would have released something like four times more energy than the famous Tunguska event and could have caused considerable damage. RAPTOR could detect an asteroid of that size out to the distance of the moon and could issue an alert to trigger real time follow-up observations.  RAPTOR's fast cadence imaging generating 100 wide-field stereoscopic images per hour will allows us to track the trajectory in 3-D. That capablility, when coupled with our real time alert system, makes RAPTOR well suited for the detection and study of Earth grazing objects and killer asteroids.

\subsection{Flares on Solar Type Stars}

There is some plausible evidence that suggests that solar type stars can occasionally generate giant flares, with durations of seconds to hours, that are as much as 10 million times more powerful than the largest ever detected from the sun \cite{Schaefer89}. Solar type stars are slow rotators and should not have sufficient free energy reservoirs to generate such enormous flares. Planetary systems around those normal, isolated, F8-G8 spectral-type main sequence stars are often thought to be the best incubators for intelligent life. If real, the giant flares would have a profound impact on life in those solar systems. It has been suggested that the giant flares could originate in systems that are similar to RS CVn magnetic binaries, but with the companion scaled down from a star to a Jovian-type planet.\cite{Rubenstein00} Unfortunately, prompt follow-up  observations of one of these giant flare candidates have never been made, so their precise nature is unknown. The problem is that typical astronomical observations are of a narrow field, employ long exposures that wash out short duration signals, and are not analyzed in real time. As a consequence, the frequency of giant flares from solar type stars, as well as their very existence is not well established. RAPTOR's combination of wide field imaging at fast cadence, real time transient recognition, and follow-up is ideal for exploring the reality and properties of the giant flares.

\subsection{Extra-Solar Planets}

The search for extra-solar planets is another area where RAPTOR has discovery potential. Transits of the parent star by a large planet can sometimes be measured by telescopes of modest aperture.\cite{Charbonneau99} In fact, the first planetary transient was discovered by a small telescope with an aperture of only 4 inches in diameter. Since then, about 60 stars in the galactic disk have been found to undergo periodic eclipses by small, dark bodies, many of which are likely to be planets.\cite{Udalski02} The key to these searches is the precision photometry made possible by fast cadence imaging. That fast cadence permits binning and precise detrending which enables detection of the weak modulation signal.  The fast cadence of the RAPTOR sky monitoring system and its generation of simultaneous image pairs will help suppress systematic photometry errors and allow a broad area search for planetary transits. 

\subsection{Mapping the Galaxy in 3-D}

The RAPTOR-P system, which will nightly monitor 10,000 square-degrees in two colors down to $\sim 16^{th}$ magnitude, will be a powerful tool for discovering new variable stars. New RS CVn binaries, contact binaries, Algols, and pulsating stars will be discovered.\cite{Paczynski01a} This two-color variability study will re-invigorate a broad range of inquiry in traditional stellar and galactic astronomy.

 One example is the study of the RR Lyrae stars. RR Lyrae are old stars whose distribution provides clues about the formation of the Galaxy. They make good candles for estimating distances and their large amplitude of variation make them easy to identify. As such, they been have used to study the galactic bulge and the outer halo. But the sample of the brighter RR Lyrae that are within a few parsecs is biased and incomplete because a systematic modern survey over a substantial fraction of the sky has not been available.\cite{Amrose01} The main source of RR Lyrae stars covering wide areas of the sky is
still photographic surveys. RAPTOR will discover and measure several thousand RRab and RRc stars, enough to trace the spatial density of the thick disk and the inner halo, and provide an important connection between surveys of the outer and inner galaxy. 

\subsection{Nearby Supernovae and Cosmology}

Type Ia supernovae (SNe Ia) are used as standard candles for studying very distant objects and the global properties of the Universe.\cite{Riess96} However, to do precision cosmology with them, one needs a deep understanding of the systematics associated with using the SNe Ia lightcurves and spectra to normalize the measurements to a standard candle. To understand these systematics we need a larger sample of nearby supernova that can be studied in detail\cite{Riess98}. But nearby SNe Ia are rare, so one needs to continually scan a large fraction of the sky and at a fast cadence to catch their lightcurves in the rising phase.  Raptor will nightly monitor a significant fraction of the sky with a sensitivity sufficient to detect SNe Ia out to a distance of 100 Mpc. We estimate that the RAPTOR-P system currently under construction should detect about 5--10 SNe Ia per year.   Those bright local events will be suitable for extensive follow-up observations using the world's best instruments at gamma-ray energies through the radio frequencies.

\section{Summary}
RAPTOR is designed to explore the dynamic optical sky for transients with durations as short as one minute. It is the first robotic optical telescope that autonomously finds and follows up on optical transients in real time. The core of the system is two wide-field telescope arrays, separated by 38 kilometers, that spectroscopically view the same 1500 square-degree piece of sky. The stereoscopic viewing, when coupled with a fast analysis pipeline, allows the system to reject false positives and robustly identify real celestial transients in real time. It also allows the system to track near earth objects that are closer than one quarter of the way to the moon in 3D. Having identified a real celestial transient, the system generates an alert and begins follow-up observations with higher spatial resolution fovea cameras and a low spectral resolution spectrograph. This integrated system will allow RAPTOR to explore a new region of discovery space in optical astronomy by mining the sky in real time. 


\acknowledgments     
 
Internal Laboratory Directed Research and Development funding supports the RAPTOR project at Los Alamos National Laboratory under DoE Contract W-7405-ENG-36.  


\end{document}